\documentclass[twocolumn,showpacs,preprintnumbers,amsmath,amssymb]{revtex4-1}
\usepackage{graphicx,color}
\usepackage{dcolumn}
\usepackage{bm}

\begin{document} 
                 
\preprint{}

\title{ The Refractive Index of Silicon at $\gamma$ Ray Energies.}
                  
\author{D.~Habs$^{1,2}$, M.M.~G\"unther$^2$, M.~Jentschel$^3$ and W. Urban$^3$}
\affiliation{$^1$Ludwig-Maximilians-Universit\"at M\"unchen, 
D-85748 Garching, Germany \\
$^2$ Max-Planck-Institut f\"ur Quantenoptik,  D-85748 Garching, Germany \\
$^3$ Institut Laue-Langevin,  F-38042 Grenoble, France}

\date{\today}
\begin{abstract}
For X-rays the real part of the refractive index, dominated by Rayleigh scattering, is
negative and converges to zero for higher energies. For $\gamma$ rays a positive component, related
to Delbr\"uck scattering, increases with energy and becomes dominating.
The deflection of a monochromatic $\gamma$ beam due to refraction was measured by
placing a Si wedge into a flat double crystal spectrometer. Data were obtained in an
energy range from 0.18 - 2 MeV. The data are compared to theory, taking into account
elastic and inelastic Delbr\"uck scattering as well as recent results on the energy
dependence of the pair creation cross section. Probably a new field of $\gamma$ optics
with many new applications opens up.
\end{abstract}


\maketitle

 In optics the index of refraction $n(E_{\gamma})=1+\delta(E_{\gamma})+i\beta(E_{\gamma})$ 
is split into a real part $\delta$ and an imaginary part $\beta$, 
describing refraction and absorption, respectively. 
In this publication we report on the first
measurement of $\delta$ for silicon up to 2 MeV and the totally unexpected
finding that it changes sign above 0.7 MeV, causing $n$ to be larger than 1. The
index deviation $\delta$ is only $10^{-9}$ and is explained by 
Delbr\"uck scattering and virtual pair creation in the high
nuclear electric field. We find that higher order Delbr\"uck scattering 
gives leading contributions to $\delta$. Thus, extrapolating 
our results  towards high Z atoms like gold, we expect 
a much larger $\delta$ in the range of about $10^{-5}$, opening up  
a totally new field of refractive $\gamma$ optics.

The use of refractive optics in combination with diffracting elements has been a fast developing field for X-rays 
up to 200 keV \cite{nielsen01} and has been used   
for focusing to a small spot size ($\mu$m - 10 nm) \cite{snigirev11}.
The refractive real part can be  directly calculated by Kramers-Kronig dispersion relations from the absorptive cross sections.
For X-ray lenses it is determined by the virtual photo effect 
(Rayleigh scattering) and follows the law \cite{lengeler99}:

\begin{equation}
  \label{eq:real}
 \delta_{photo}=-2.70 \cdot \frac{\lambda^2\cdot \rho \cdot Z}{A}\cdot 10^{-6}
       \propto \frac{1}{E_{\gamma}^2} \mbox{.}
\end{equation}

Here the photon wave length $\lambda$ is measured  in 
$\AA$, $\rho$ in $g/cm^3$, $Z$ is the atomic number and $A$ is the atomic 
mass in g. Obviously, $\delta_{photo}$ converges with $1/E_{\gamma}^2$ very 
fast towards zero, limiting significantly the construction of focusing optics 
at higher energies. A reasonable focal length $f$ could be obtained by a 
large number of lenses $N$,  since $ f= R/(2\cdot\delta\cdot N)$, where $R$ is the 
radius at the apex of the parabolically shaped, concave lenses. In case of 
hard X-rays several hundreds lenses are used. For MeV $\gamma$ rays this 
number $N$ would become extremely large and absorption will prevent their use.
The argumentation bases on the extrapolation of the $1/E_\gamma^2$ scaling of the virtual photo effect, for 
 which experimental knowledge is only available up to about 200 keV.

\begin{figure} [h]
   \begin{center}
   \begin{tabular}{c}
   \includegraphics[height=5.3cm]{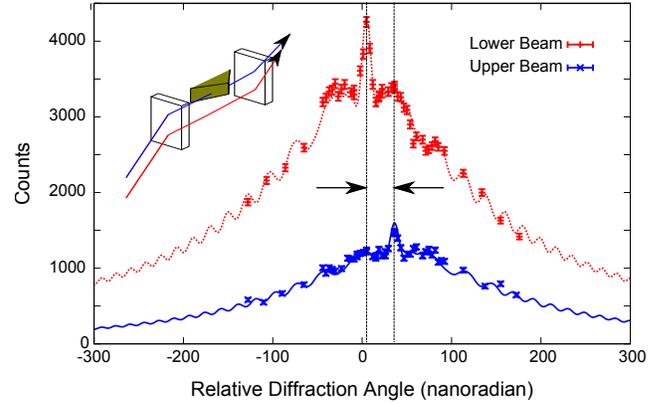}
   \end{tabular}
   \end{center}
   \caption[lens] 
   { \label{fig:exp} 
Illustration of the measurement principle. The $\gamma$ beam is coming from
the ILL high-flux reactor. A silicon wedge placed between the two crystals of
the GAMS5 spectrometer deflects only the upper part of the beam, while the lower beam is propagating in air. The two line shapes were taken during preparation for illustration purposes with the 184 keV line of $^{168}$Er with a very long acquisition time. During the actual measurement, the acquisition time was about ten times shorter. The upper (blue) beam shows a reduced intensity compared to the lower (red) beam due to  absorption in the wedge. 
The spectra show ``pendel solutions'' like intensity oscillations typical for a 
double crystal spectrometer \cite{kessler01}. One clearly sees the angular 
deflection by the wedge.}
   \end{figure} 
   
An experiment following the concept of \cite{deutsch84}, but using $\gamma$ rays from 
an in-pile target at the neutron high flux reactor at ILL in Grenoble was carried out.
Samples with a total mass of about 10 g and a total surface of about 30 cm$^2$ are 
placed in a thermal neutron flux of $4 \times 10^{15}/(s~cm^2)$ and yielding emission rates for $\gamma$ energies of up to 10$^{15}$ s$^{-1}$. For the present experiment we use 
very intense transitions of $^{36}$Cl and $^{156,158}$Gd (and during a preparation stage, of  $^{168}$Er).
The beam from the sample is collimated over a total distance of 17 meters with a cross 
section of $4\times20$ mm$^2$. Behind the spectrometer, a movable three meter long collimation 
system separates the diffracted beam from the direct beam. The detection is done via 
a calibrated HP-Ge detector. A schematic layout of the experimental principle is shown in Fig. \ref{fig:exp}.

The spectrometer is equipped with 2.5 mm thick single crystals of silicon, the lattice spacing of which has been indirectly measured with $\Delta d/d \simeq 5 \times 10^{-8}$ 
(for details see \cite{dewey}). The angular acceptance width of the crystals at $\gamma$-ray energies  is in the order of nanoradian. The spectrometer is set in non-dispersive 
geometry, where the second crystal rocks around the parallel position with respect 
to the first one. In this geometry the first crystal diffracts an energy band defined 
by the divergence of the incoming beam (a few microradian). 
In this geometry the instrument does not act as high resolution monochromator. 
The line shapes are obtained via summing counts in a 2 keV wide window of the Ge spectrum, taken for 
each crystal position. This discriminates background $\gamma$ rays or unwanted diffraction orders via the energy resolution of the detector. The advantage of the non-dispersive geometry is its high sensitivity to beam deflection between the crystals induced by a refractive prism (see below). At the same time systematic errors such due to  Doppler-broadening of the $\gamma$ rays (thermal motion, recoil motion) 
and vertical divergence \cite{Bearden} are at minimum. 

\begin{figure} 
   \begin{center}
   \begin{tabular}{c}
   \includegraphics[height=6cm]{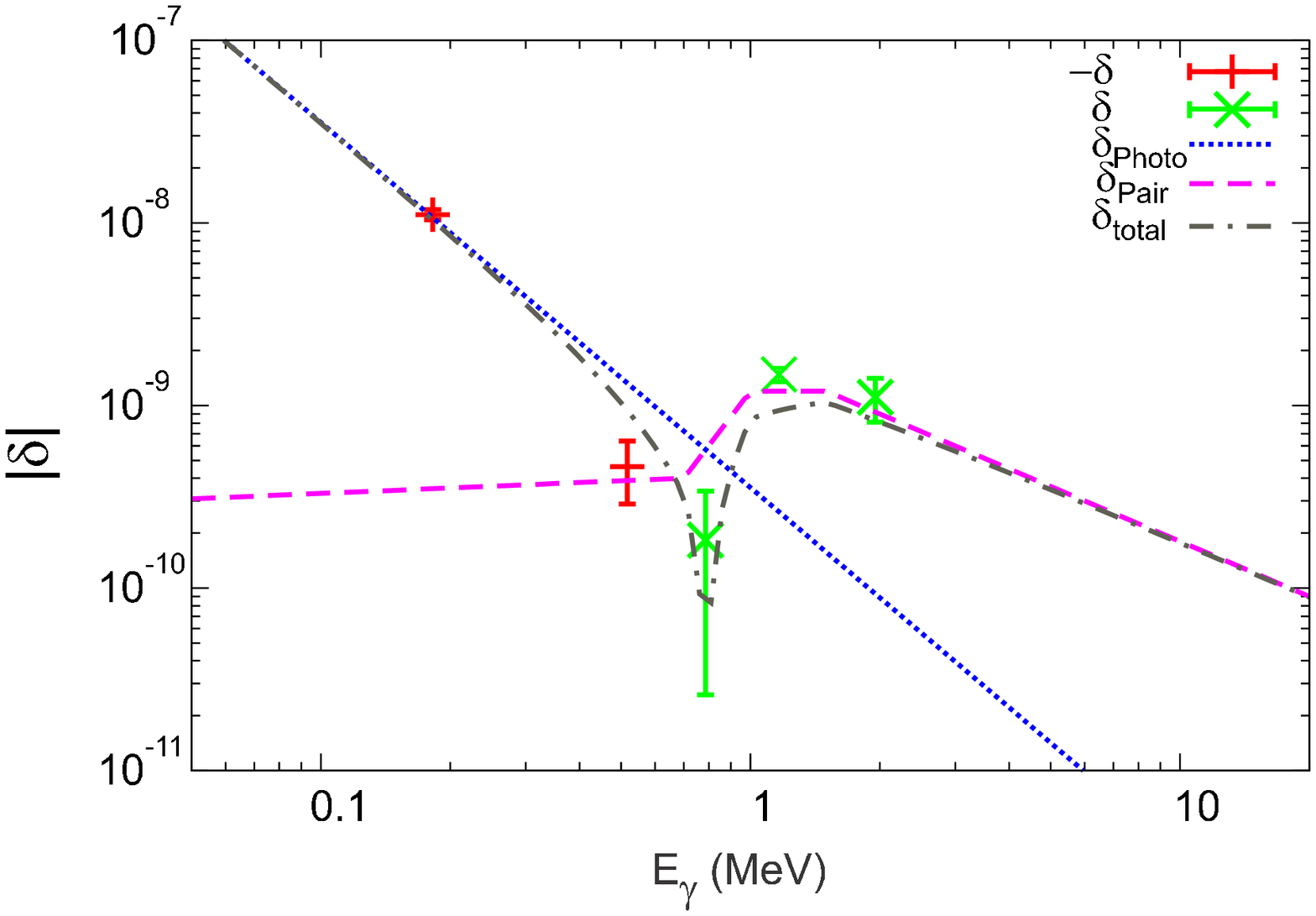}
   \end{tabular}
   \end{center}
   \caption[lens] 
   { \label{fig:delta} Newly measured index of refraction $|\delta|$ for
$\gamma$ energies up to 2 MeV. The blue dashed curve shows the negative
$\delta$ from the virtual photo effect, which is confirmed by the measured
values for lower $\gamma$ energies. For the positive $\delta$ of virtual 
pair creation we used a shape from our dispersion relation calculations. 
Also the superposition of the two $\delta$ contributions is shown.}
   \end{figure}
 
\begin{table}
\caption{$\gamma$ energies with measured index of refraction $\delta$
for silicon using $^{36}$Cl and $^{156,158}$Gd sources.}
\smallskip
\begin{center}
\begin{tabular}{|c|c|c|c|} \hline

$E_{\gamma}$ 	& wedge           	& background	& Isotope 	\\ 
   keV          				& $\delta$           &  $\delta$		&					\\ \hline 
517 							& (-4.63 $\pm 1.75)\cdot 10^{-10}$	&	&  $^{36}$Cl\\
786 							&(+1.83 $\pm 1.57)\cdot 10^{-10}$	&	&  $^{36}$Cl\\
1165						&(+1.48$\pm 0.13 )\cdot 10^{-9}$ 	& (+0.32$\pm 0.01)\cdot 10^{-9}$&  $^{36}$Cl\\
1951						&(+1.11$\pm 0.30)\cdot 10^{-9}$  	&&  $^{36}$Cl\\
182 							&(-1.11$\pm 0.07)\cdot 10^{-8}$		 &(-0.67$\pm 0.01)\cdot 10^{-9}$&   $^{158}$Gd\\ \hline
\end{tabular}
\end{center}
\end{table}

Between the two crystals a silicon prism with an edge angle of 160 degrees and optically polished flat 
faces was mounted. The upper half of the $\gamma$ beam was passing through the wedge, while the lower 
half passes underneath through air. The second crystal is supposed to compare angular deviations of 
the upper and lower beam with a sensitivity of $10^{-9}$ rad. The switching between the two halves 
of the beam is realized via a lead shutter, which is installed on the collimation system behind 
the spectrometer. This assures that the movement of the lead shutter is mechanically decoupled from the 
spectrometer. The acquisition time was kept as short as possible to minimize drift problems. For each energy about 30 pairs of 
scans were taken and averaged. The temperature, pressure and humidity at the spectrometer were monitored and time variations of the lattice spacing and the refractive index of air were corrected as described in \cite{dewey}.
For background measurements we compared both beams, removing the wedge from the spectrometer. However, this was only carried out once at the end of the measurement, since it represents a serious intervention perturbing the stability of the instrument. The background values were not subtracted from the data, since only a few are available. Further, a direct conversion of the measured deviation angles into absolute values of $\delta$ is difficult, since it requires an exact knowledge of the absolute alignment of the wedge with respect to the beam. Since we are interested in the energy dependence, we have used the result at 182 keV for normalization to equation (\ref{eq:real}).
The measured values of the index of refraction are compiled in Table 1 and shown in Fig.  \ref{fig:delta}.

\begin{figure*}[t!]
   \begin{center}
   \begin{tabular}{c}
   \includegraphics[height=7cm]{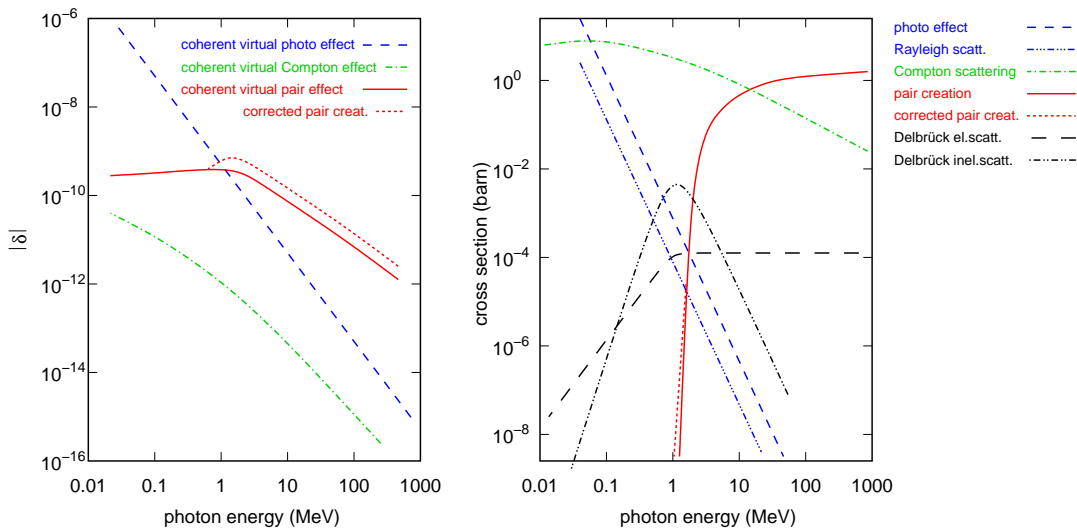}
   \end{tabular}
   \end{center}
   \caption[lens] 
   { \label{fig:theoold} 
Theoretical absolute real part of the index of refraction $\delta$ (left) 
and the absorption cross sections (right) in silicon as a function 
of $E_{\gamma}$ from \cite{hubbell80}. The three contributions 
to $\delta$ from the virtual photo effect, virtual Compton effect and 
the virtual pair creation were deduced via the 
Kramers-Kronig dispersion relations from the corresponding absorption cross 
sections. The dotted red line (right) shows a recently measured pair 
creation cross section close to the threshold \cite{jentschel11}. 
Its enhancement over the Bethe-Heitler prediction can again be explained 
by dispersion relations. We also show the cross section for Rayleigh, 
elastic Delbr\"uck (virtual pair creation, recoilless) and inelastic 
Delbr\"uck (virtual pair creation, momentum transfer to the nucleus) 
scattering. The important new finding is the dominance of the positive 
$\delta$ due to virtual pair creation above about 1 MeV $\gamma$ rays.}
   \end{figure*} 

The measurement bases on  a
homogeneous lattice spacing.  The maximum variation of the lattice spacing 
can be estimated by considering the measured width of the rocking 
curve and comparing it to calculations of dynamical diffraction theory of 
ideal crystals. Such evaluations are sensitive for $\gamma$ ray energies above 700 keV, 
since the rocking curve is becoming sufficiently narrow. Experimentally, a deviation from theory in the order of 30 nanoradian is observed. It is almost completely related to vibrations of the crystals of about 20 nanoradian, each.
A relative variation of the lattice spacing of $\Delta d/d < 10^{-6}$ could be estimated. A slight energy dependence of the background 
measurement can be seen. At present we interpret this effect by residual 
surface tensions on the crystal. In \cite{mana} theoretical investigations 
were carried out indicating that the diffraction angle in Laue geometry is 
defined within a surface layer. Since, the depth of this layer is increasing with increasing energy, the 
low energy measurements are more sensitive.
   
An interpretation of the experimental results is possible by demonstrating
that the process of virtual pair creation (Delbr\"uck scattering) takes over 
at higher $\gamma$ energies with a $\delta_{pair}$, leading to a sign change 
of the refractive index. In Delbr\"uck scattering the $\gamma$ ray interacts
with the strong electrical field of the nucleus, causes virtual pair creation
and then is re-scattered. Averaging over a sphere with the diameter of 
the Compton wave length of the electron (400 fm), we obtain an electric nuclear 
Coulomb field for silicon of about the Schwinger field of $E_s=1.2\cdot 10^{18}$V/m,
pointing to a situation where perturbative first-order QED is not sufficient.
The dominance of $\delta_{pair}$ for higher $\gamma$ 
energies can be explained by a refined use of dispersion relations 
\cite{bjorken65,jackson75,toll52}, which remain valid even for non-perturbative
QED.

In the right hand side of Fig. \ref{fig:theoold}, we show  
contributions to the absorption cross sections for $\gamma$ rays 
\cite{hubbell80} together with cross sections of different scattering processes. We divided the 
important Delbr\"uck scattering into an elastic part, where in a 
M\"ossbauer-like recoilless scattering 
the recoil is taken up by the total crystal $\sigma_{Delb,elas.}$ 
and an inelastic cross section $\sigma_{Delb,inelas.}$, where 
the scattered $\gamma$ quantum has lost the recoil energy 
to the nucleus. We roughly calculate
the angle integrated cross sections from theory \cite{rohrlich52,bethe52}
up to the M\"ossbauer limit $\Theta_{elas}$ for the
elastic Delbr\"uck scattering $\sigma_{Delb,elas.}$ and beyond $\Theta_{elas}$
for the inelastic Delbr\"uck scattering $\sigma_{Del,inels.}$. Both are shown
in Fig. \ref{fig:theoold} .

We now use the dispersion relations \cite{jackson75,nielsen01,toll52} to 
calculate the real parts $\delta$ of the refraction index on the left side 
of Fig. \ref{fig:theoold}. We introduce the outgoing complex forward scattering 
amplitude $A_{f}(E_{\gamma})=A_{rf}(E_{\gamma}) + iA_{if}(E_{\gamma})$,
with the real part $A_{rf}(E_{\gamma})$ and the imaginary part 
$A_{if}(E_{\gamma})$. 
The virtual processes with coherent cross sections $\sigma_{sca}$
contribute to the real part $A_{rf}$, the real processes with cross sections
$\sigma_{abs}$, but also inelastic virtual processes contribute to the imaginary 
part $A_{if}$ of the coherent scattering amplitude
 \begin{equation}
  \label{eq:kramer0b}
A_{rf}(E_{\gamma})=\frac{E_{\gamma}^2}{\pi(4\pi^2 \hbar c)^2}
\lim_{\epsilon\to 0+}\int^\infty_0
\frac{A_{if}(E) dE}{E(E^2-(E_{\gamma}+i\epsilon)^2)} \mbox{.}
\end{equation}

By the optical theorem \cite{jackson75,bjorken65},
the imaginary part of the forward-scattering amplitude
is related to the total absorption cross section, 
$\sigma_{abs}(E_{\gamma})= 2\cdot \lambda \cdot A_{if}(E_{\gamma})$, 
or, $ A_{if}(E)=(\frac{E}{4\pi \hbar c}) \cdot \sigma_{abs}(E)$.
We thus obtain
 \begin{equation}
  \label{eq:kramer0c}
A_{rf}(E_{\gamma})=\frac{E_{\gamma}^2}{2\pi^2 \hbar c}
\lim_{\epsilon\to 0+}\int^\infty_0
\frac{\sigma_{abs}(E) dE}{E^2-(E_{\gamma}+i\epsilon)^2} \mbox{.}
\end{equation}
This relation connects the total absorption cross section $ \sigma_{abs}(E)$
with the forward coherent scattering cross section
 \begin{equation}
  \label{eq:kramer0c}
\frac{d \sigma_{sca}}{d \Omega}(forward)= |A_{rf}(E_{\gamma})|^2 \mbox{.}
\end{equation}
$A_{rf}(E_{\gamma})$ is related to the real part of the index of refraction 
$\delta(E_{\gamma})$
\begin{equation}
  \label{eq:kramer1}
\delta(E_{\gamma})= 
\frac{\lambda^2}{2\pi}\cdot N_c \cdot A_{rf}(E_{\gamma}) \mbox{,}
\end{equation}
where $N_c$ is the number of nuclear scattering centers per volume. 
With these formulas we have calculated the real
parts $\delta_{photo}, \delta_{Compton}$ and $\delta_{pair}$
for the  corresponding three processes photo effect, Compton effect and pair 
creation.

Fig. \ref{fig:theoold} shows these three components together with the experimentally 
measured $\delta_{photo}$  from Eq. (1). The strongly rising pair creation 
cross section and inelastic Delbr\"uck scattering 
results in a positive $\delta_{pair}$, while the decreasing cross sections of 
the photo effect and Compton scattering result in a negative $\delta_{photo}$ 
and a negative $\delta_{Compton}$. The surprising result is, that  
$|\delta_{pair}|$ becomes larger than  $|\delta_{photo}|$, which we wanted 
to verify experimentally, including the sign change of $\delta$.

Historically, J.S.~Toll and J.A.~Wheeler \cite{toll52} predicted wrongly 
in 1952 for lead (Z=82) and 1 MeV $\gamma$ quanta that $|\delta_{pair}|$
should be about a factor of $10^3$ smaller than $|\delta_{photo}|$, discouraging
experimentalists from searching for a $\delta_{pair}$ contribution. Via   
dispersion relations they calculated $\delta_{pair}$ in first order from 
the absorptive Bethe-Heitler cross section. When we included also the 
inelastic, higher-order Delbr\"uck scattering, we obtained the much larger
$\delta_{pair}$ (red curve), reaching the up-shift of $\delta_{pair}$ for 
1 MeV to $1\cdot 10^{-9}$, the range of the measured values.
 A further increase of $\delta_{pair}$ and 
$A_{rf, pair}(E_{\gamma})$ (dashed red curve) occurs, because 
experimentally a strong enhancement of pair creation close to threshold over 
the predicted Bethe-Heitler cross section \cite{jentschel11} is observed.
Since the contribution to $\delta$
in Eq. (3) becomes very large if $E$ is close to $E_{\gamma}$,
and if $\sigma_{abs}$ is steeply sloped as a function of $E$,
this strong increase of the absorption cross section close to $2\cdot m_ec^2$
 increases the value of $\delta_{pair}$ significantly.  
Thus, the newly measured absolute values of $\delta_{pair}$ and 
$\delta_{photo}$ can be explained naturally, using the dispersion relations.
The higher order terms with $Z^4$ and $Z^6$ contribute much stronger 
to $\delta_{pair}$ compared to the first order term with $Z^2$,  showing that
this corresponds to non-perturbative high field QED, which is addressable by
dispersion relations.  The value of $\delta_{pair}$ is  rather constant  at 
low energies, then rises to a maximum  between 1 MeV and 2 MeV,
followed by a fall-off with $E_{\gamma}^{-1}$ for higher $\gamma$ energies. 
The  derivation of  Kramers-Kronig dispersion relations bases on a sum decomposition 
$\delta_{pair}(E_{\gamma},Z,A,\rho)=\frac{\rho}{A}\sum_{n=1}^\infty
c_n(E_{\gamma})\cdot Z^{2n}$ assuming a fast convergence of the coefficients $c_n$.
In case of high fields the convergence assumption is not valid. Although a strict  
theory does not yet exist one can expect that the $Z^4$ term due to the 
Delbr\"uck cross section in the dispersion integral and
higher $Z^{2n}$ terms become dominant. 
The present experiment is focusing 
on silicon as for this material technologies for micro-lithographic
lens manufacturing are elaborated \cite{schroer10}. However, future experiments will focus 
on high Z materials like gold (Z=79). A naive scaling with $(79/14)^6$
results in a $\delta_{pair}$ of $3\cdot 10^{-5}$ in the 1 MeV range. At present
the development of small biconvex 1 mm diameter gold $\gamma$ lenses is underway. We expect
a focal length of 3 m and want to determine $\delta_{pair}$ accurately by 
measuring the focal length. As $\delta_{pair}$ increases much stronger with Z 
compared to $\delta_{photo}$, the switch-over energy from positive to negative
$\delta$ is shifted to much lower energies.
As shown in Fig. \ref{fig:delta},  we have a destructive superposition of 
both contributions of $\delta_{photo}$ and $\delta_{pair}$ at the switch-over 
energy of about 700 keV for silicon and $\delta$ is positive for 
higher $\gamma$ energies. For positive $\delta$ convex lenses are required
for focusing. 

We will develop a whole new $\gamma$ optics tool box: $\gamma$ 
lenses for focusing, gold prisms for deflection, short $\gamma$ wave guides
from gold with total internal reflection. With a combination of refractive
and diffractive $\gamma$ optics, we can realize very efficient $\gamma$
monochromatisation down to a band width of $10^{-6}$, allowing to address 
individual nuclear levels up to the neutron binding energy \cite{habs12}.
With the upcoming, $10^7$ times more brilliant and intense $\gamma$ beams
of MEGa-Ray (Livermoore, USA, 2013) \cite{barty11} and ELI-NP (Bucharest,
Romania, 2015) \cite{ELI-NP11}, compared to the present worldwide best
$\gamma$ facility HI$\gamma$S (Duke University, USA), a broad field of
new applications with nuclear resonance excitation (radioactive waste
management, imaging of $^7$Li in batteries for green energy, production
of about 50 new medical radioisotopes with high specific activity for
diagnostics and therapy \cite{habsa11} etc..) opens up in nuclear 
photonics \cite{habs12}. Also high-resolution nuclear spectroscopy
with a few eV resolution \cite{thirolf12} and efficient 
$\gamma$ astronomy will flourish.


\end{document}